# Nonlinear Model Predictive Control for Enhanced Path Tracking and Autonomous Drifting Through Direct Yaw Moment Control and Rear-Wheel-Steering

Gaetano Tavolo[1](✉), Pietro Stano[1], Davide Tavernini[1], Umberto Montanaro[1], Manuela Tufo[2,3], Giovanni Fiengo[2,3], Pietro Perlo[4], and Aldo Sorniotti[5]

[1] University of Surrey, Guildford GU2 7XH, UK
g.tavolo@surrey.ac.uk
[2] University of Sannio, 82100 Benevento, Italy
[3] Kineton R&D, 80146 Naples, Italy
[4] Interactive Fully Electrical VehicleS (IFEVS), 10040 Turin, Italy
[5] Politecnico di Torino, 10129 Turin, Italy

**Abstract.** Path tracking (PT) controllers capable of replicating race driving techniques, such as drifting beyond the limits of handling, have the potential of enhancing active safety in critical conditions. This paper presents a nonlinear model predictive control (NMPC) approach that integrates multiple actuation methods, namely four-wheel-steering, longitudinal tyre force distribution, and direct yaw moment control, to execute drifting when this is beneficial for PT in emergency scenarios. Simulation results of challenging manoeuvres, based on an experimentally validated vehicle model, highlight the substantial PT performance improvements brought by: i) vehicle operation outside the envelope enforced by the current generation of stability controllers; and ii) the integrated control of multiple actuators.

## 1 Introduction

Since 2014, the EU has mandated vehicle stability controllers (VSCs), enhancing active safety during emergency manoeuvres, through the limitation of the yaw rate error, sideslip angle, and longitudinal tyre slip [1, 2]. While these systems are effective in supporting the average human driver, they may be overly conservative for highly automated vehicles (AVs) [3]. In parallel, powertrain electrification and active chassis control systems offer new AV control opportunities [4]. A current trend in AV research involves

This work was supported by the Horizon 2020 Programme of the European Commission under Grant 101006953 (Multi-Moby project).

emulating expert driving techniques, such as drifting, by using path tracking (PT) controllers that push the vehicle beyond the conventional VSC-related stability constraints, potentially improving road safety [5]. Several PT algorithms from the literature, e.g., [6, 7], achieve controlled drifting by tracking dedicated sideslip angle and yaw rate profiles, in addition to the reference trajectory. However, such controllers are not designed to induce drifting only when necessary to track a challenging trajectory, and are often demonstrated in scenarios with negligible vehicle speed variations or increasing speed, along circular paths or during drift parking, typically with rear-wheel-drive AVs [8, 9]. On the contrary, typical real-world emergency manoeuvres involve significant speed reductions through braking, imposed either by the human/automated driver and/or the VSC. Although the recent studies in [5] and [10] demonstrate the safety benefits of autonomous drifting in realistic scenarios, the literature lacks a performance assessment of different chassis actuation suites, and especially rear-wheel-steering (RWS), in terms of accident avoidance through AV control beyond the VSC-related boundaries. This paper targets the gap by proposing an NMPC algorithm that integrates four-wheel-steering, longitudinal tyre force distribution, and direct yaw moment (DYM) actuation. Simulation results with a high-fidelity model along two challenging manoeuvres show the controller's capability to perform drifting, and the benefit of rear steering actuation for achieving tighter cornering.

## 2 Control Architecture

Three alternative NMPC PT formulations are considered: i) $NMPC_{M_z,\delta_r}$, which is the novelty of the study, and controls (independently from each other) the time derivatives of the front and rear steering angles, $\dot{\delta}_f$ and $\dot{\delta}_r$; the time derivative of the longitudinal tyre force on the front axle, $\dot{F}_{x,f}$; the front-to-total force distribution factor in braking, $p_b$; and the time derivative of the DYM, $\dot{M}_z$; ; ii) $NMPC_{M_z}$, which, compared to $NMPC_{M_z,\delta_r}$, excludes $\dot{\delta}_r$ control; iii) $NMPC_{bas}$, which, compared to $NMPC_{M_z}$, excludes $\dot{M}_z$ and $p_b$ control. The control allocation (CA) algorithm is detailed in [5], and is integrated with a rule-based VSC, including a PID anti-lock braking system (ABS). In the $NMPC_{M_z,\delta_r}$ and $NMPC_{M_z}$ simulations, the VSC intervention thresholds are relaxed, to allow operation beyond the limits of handling.

### 2.1 Prediction Model Formulations

The prediction models are based on the single-track formulation in [5]. For brevity, only the updated longitudinal and lateral force balance and yaw moment balance equations for the $NMPC_{M_z,\delta_r}$ case are reported:

$$
\begin{aligned}
\dot{v}_x = \frac{1}{m}[F_{x,f}\cos(\delta_f) - F_{y,f}\sin(\delta_f) + F_{x,R}\cos(\delta_r) \\
- F_{y,r}\sin(\delta_r) - F_{x,Mz} - F_{drag} - F_{roll} + mv_y\dot{\psi}] \\
\dot{v}_y = \frac{1}{m}[F_{x,f}\sin(\delta_f) + F_{y,f}\cos(\delta_f) \\
+ F_{x,r}\sin(\delta_r) + F_{y,r}\cos(\delta_r) - mv_x\dot{\psi}]
\end{aligned}
$$

$$\ddot{\psi} = \frac{1}{I_z}[F_{x,f}\sin(\delta_f)l_f + F_{y,f}\cos(\delta_f)l_f - F_{y,r}\cos(\delta_r)l_r - F_{x,r}\sin(\delta_r)l_r + M_z] \quad (1)$$

where $m$ is the vehicle mass; $I_z$ is the yaw mass moment of inertia; $l_f$ and $l_r$ are the front and rear semi-wheelbases; $F_{drag}$ and $F_{roll}$ are the aerodynamic drag and rolling resistance force; $F_{x,i}$ and $F_{y,i}$ are the longitudinal and lateral tyre forces of the axle $i$, where the subscript $i = f, r$ refers to the front or rear axles; and $M_z$ is the direct yaw moment generated through the actuation of the friction brakes, while $F_{x,M_z}$ is the corresponding longitudinal tyre force contribution. The lateral axle forces, $F_{y,i,}$, are calculated through a simplified version of the Pacejka magic formula, whose vertical tyre force inputs account for both the static contribution and the longitudinal load transfer, based on the measured longitudinal acceleration, $a_{x,meas}$.

### 2.2 Optimal Control Problem

At each time step, the NMPC algorithm computes an optimal control sequence that minimizes a cost function, see its discrete form in [5], based on the outputs from the prediction models, which are expressed through the following continuous time formulation:

$$\dot{x}(t) = f(x(t), u(t), w(t)) \quad (2)$$

where $x$ is the state vector, and $u$ is the control input vector, which, for the three considered controller configurations, are expressed as:

$$\begin{aligned} x_{M_z,\delta_r} &= \left[v_x, v_y, \dot{\psi}, s, e_y, e_\psi, \delta_f, F_{x,f}, M_z, \delta_r\right] & u_{M_z,\delta_r} &= \left[\dot{\delta}_f, \dot{F}_{x,f}, p_b, \dot{M}_z, \varepsilon_{Mz}, \dot{\delta}_r\right] \\ x_{M_z} &= \left[v_x, v_y, \dot{\psi}, s, e_y, e_\psi, \delta_f, F_{x,f}, M_z\right] & u_{M_z} &= \left[\dot{\delta}_f, \dot{F}_{x,f}, p_b, \dot{M}_z, \varepsilon_{Mz}\right] \\ x_{bas} &= \left[v_x, v_y, \dot{\psi}, s, e_y, e_\psi, \delta_f, F_{x,f}\right] & u_{bas} &= \left[\dot{\delta}_f, \dot{F}_{x,f}\right] \end{aligned} \quad (3)$$

where $s$ is the distance along the path; $e_y$ is the lateral position error; $e_\psi$ is the heading angle error; and $\varepsilon_{Mz}$ is the slack variable associated with a soft constraint on the DYM. The online data vector, $w$, is the same for $NMPC_{M_z,\delta_r}$ and $NMPC_{M_z}$, i.e., $w_{M_z,\delta_r} = w_{M_z} = \left[a_{x,meas}, \mu, \rho_{ref}, M_{z,min}, M_{z,max}\right]$, where $\mu$ is the tyre-road friction factor, which is assumed constant along the prediction horizon ($H_p$), while the reference road curvature ($\rho_{ref}$) and the maximum and minimum DYM values ($M_{z,max}$ and $M_{z,min}$) vary according to [5]. In $NMPC_{bas}$, $w$ excludes $M_{z,min}$ and $M_{z,max}$. The output vectors and their reference values include the PT error variables and the states associated with the control inputs:

$$\begin{aligned} z_{M_z,\delta_r} &= \left[v_x, e_y, e_\psi, \delta_f, \delta_r, F_{x,f}, M_z\right] & z_{M_z,\delta_r,ref} &= \left[v_{x,ref}, 0, 0, 0, 0, 0, 0\right] \\ z_{M_z} &= \left[v_x, e_y, e_\psi, \delta_f, F_{x,f}, M_z\right] & z_{M_z,ref} &= \left[v_{x,ref}, 0, 0, 0, 0, 0\right] \\ z_{bas} &= \left[v_x, e_y, e_\psi, \delta_f, F_{x,f}\right] & z_{bas,ref} &= \left[v_{x,ref}, 0, 0, 0, 0\right] \end{aligned} \quad (4)$$

where the reference longitudinal speed, $v_{x,ref}$, is assumed to be known and variable along $H_p$. For all controllers, hard constraints are set on the steering angles, the front longitudinal tyre force, and their variation rates, as well as on the rear longitudinal tyre force:

$$\begin{aligned}
&-\delta_{i,max} \leqslant \delta_i \leqslant \delta_{i,max} \\
&-\mu_{id} F_{z,f} \leqslant F_{x,f} - F_{x,M_z} \tfrac{F_{z,f}}{F_{z,f}+F_{z,r}} \leqslant \mu_{id} F_{z,f} \\
&-\dot{\delta}_{i,max} \leqslant \dot{\delta}_i \leqslant \dot{\delta}_{i,max} \\
&\dot{F}_{x,f,min} \leqslant \dot{F}_{x,f} \leqslant \dot{F}_{x,f,max} \\
&-\mu_{id} F_{z,r} \leqslant F_{x,r} - F_{x,M_z} \tfrac{F_{z,r}}{F_{z,f}+F_{z,r}} \leqslant \mu_{id} F_{z,r}
\end{aligned} \tag{5}$$

where $\mu_{id}$ is an ideal friction factor, which marginally overestimates the real one used in the high-fidelity vehicle model to avoid underbraking, while the slip ratios are limited by the ABS, with relaxed intervention thresholds in case of $NMPC_{M_z,\delta_r}$ and $NMPC_{M_z}$; and the term $F_{x,M_z}$ is present only in $NMPC_{M_z,\delta_r}$ and $NMPC_{M_z}$.$NMPC_{M_z}$ and $NMPC_{M_z,\delta_r}$ also include a hard constraint on $p_b$, a soft constraint on $M_z$, and a hard constraint on $\dot{M}_z$:

$$\begin{gathered}
p_{b,min} \leq p_b \leq p_{b,max} \\
-\varepsilon_{Mz} + M_{z,min} \leq M_z \leq M_{z,max} + \varepsilon_{Mz}, \text{ with } \varepsilon_{Mz} \geq 0 \\
\dot{M}_{z,min} \leq \dot{M}_z \leq \dot{M}_{z,max}
\end{gathered} \tag{6}$$

where $M_{z,min}$ and $M_{z,max}$ are computed by the CA algorithm, based on the prediction of the control inputs at the previous time step.

# 3 Model Validation and Simulation Results

## 3.1 Case Study Vehicle

A four-wheel-drive electric vehicle (EV) prototype by IFEVS, with a centralized on-board electric machine per axle, is used as case study, see Fig. 1a. The EV is equipped with: i) a set of vehicle dynamics sensors, e.g., to measure the individual wheel speeds and the longitudinal and lateral velocity components; ii) an integrated GPS device with inertial measurement unit; iii) a modified commercial VSC unit to independently control the friction brake torque of each corner; and iv) a dSPACE MicroAutoBox III system for rapid control prototyping. The vehicle simulation model was implemented in the IPG CarMaker environment, and was experimentally validated along a handbrake manoeuvre, see Fig. 1b. The very good match between the experiments and the high-fidelity and prediction model results confirms: i) the reliability of the CarMaker model as a control system assessment tool; and ii) the ability of the proposed prediction models to accurately capture the system dynamics at the limit of handling.

## 3.2 Simulation Results

The NMPC implementations were evaluated along two manoeuvres, i.e., a 135-deg turn and a U-turn with braking from an initial speed of 45 km/h. For $NMPC_{M_z,\delta_r}$, different RWS angle limits, equal to 5, 10, and 15 deg, were set. Although higher than those for typical RWS systems, these values were selected to assess the potential advantages of enhanced rear steering capabilities. The 135-deg turn results are reported in Fig. 2a. The vehicle with $NMPC_{bas}$ significantly deviates from the reference trajectory, and exits the turn with unsafe lateral and orientation errors. On the contrary, the drifting behaviour induced by $NMPC_{M_z}$ and $NMPC_{M_z,\delta_r}$ enables the EV to effectively complete the manoeuvre. The DYM interventions create an asymmetric braking force distribution between the EV sides, resulting in increased yaw rate and rear slip angle magnitudes, which enhance manoeuvring agility. However, the $M_z$ profile requested by the PT controller is very different from the one generated by the vehicle through the CA algorithm, see the mismatch between the 'NMPC' and 'Plant' curves in the $M_z$ plot. This is caused by the absence of the lateral load transfer in the prediction model, and the purposely relaxed constraints on the longitudinal force and DYM, to fully utilise the tyre-road friction capability. While the $\delta_r$ actuation does not improve performance in the 135-deg turn, the counterphase actuation of $\delta_f$ and $\delta_r$ can facilitate vehicle oversteer [11], with potential benefits during more aggressive manoeuvres, such as the U-turn in Fig. 2b, for which a higher initial destabilizing effect is desirable. Since the inner wheel tends to be saturated during high lateral acceleration manoeuvring, the DYM intervention struggles generating the destabilizing effect required to induce controlled drifting, which, instead, can be more effectively induced with the RWS intervention. As a result, $NMPC_{M_z,\delta_r}$ generates higher yaw rates, reducing the DYM intervention as well as the lateral and speed tracking errors. For $NMPC_{M_z}$ and $NMPC_{M_z,\delta_r}$, Table 1 includes a set of PT key performance indicators (KPIs), including the maximum and root mean square (RMS) errors of the lateral position and speed, along with the maximum sideslip angle magnitude, indicating the extent of the controlled drifting behaviour. The RWS actuation of $NMPC_{M_z,\delta_r}$ reduces the lateral error by ~1.8 m during the U-turn test, compared to DYM intervention alone of $NMPC_{M_z}$. The best trade-off between the lateral and speed tracking errors is achieved with a maximum rear steering angle magnitude of 10 deg. A further increase to 15 deg results in higher $|\beta|_{max}$ (up to 44 deg), lower $RMS_{e_{v_x}}$, and increased $\left|e_y\right|_{max}$(by 0.17 m).

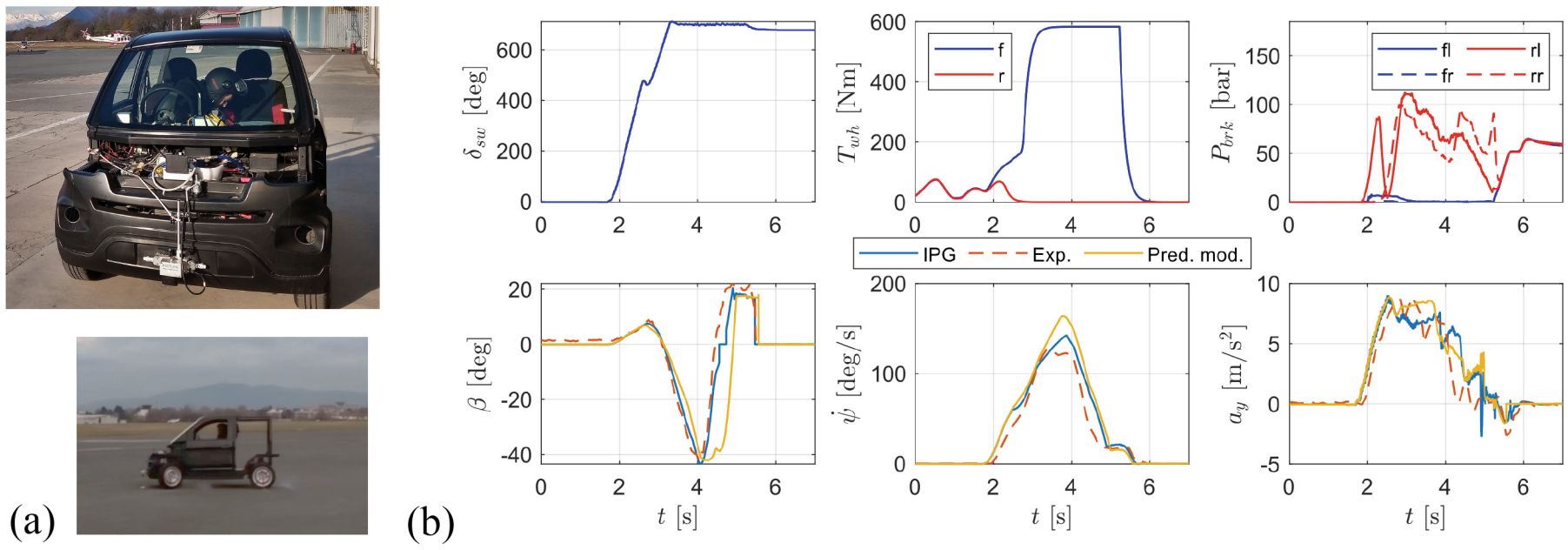

**Fig. 1.** (a) Case study EV prototype. (b) Time profiles of the hand-brake manoeuvre inputs, i.e., steering wheel angle $\delta_{sw}$, wheel torque $T_{wh}$, and braking pressures $P_{brk}$, along with the associated sideslip angle $\beta$, yaw rate $\dot{\psi}$, and lateral acceleration $a_y$. IPG: CarMaker simulation model results; Exp.: experimental results; Pred. Mod.: prediction model results. The notations 'fl', 'fr', 'rl' and 'rr' refer to the front left, front right, rear left, and rear right corners.

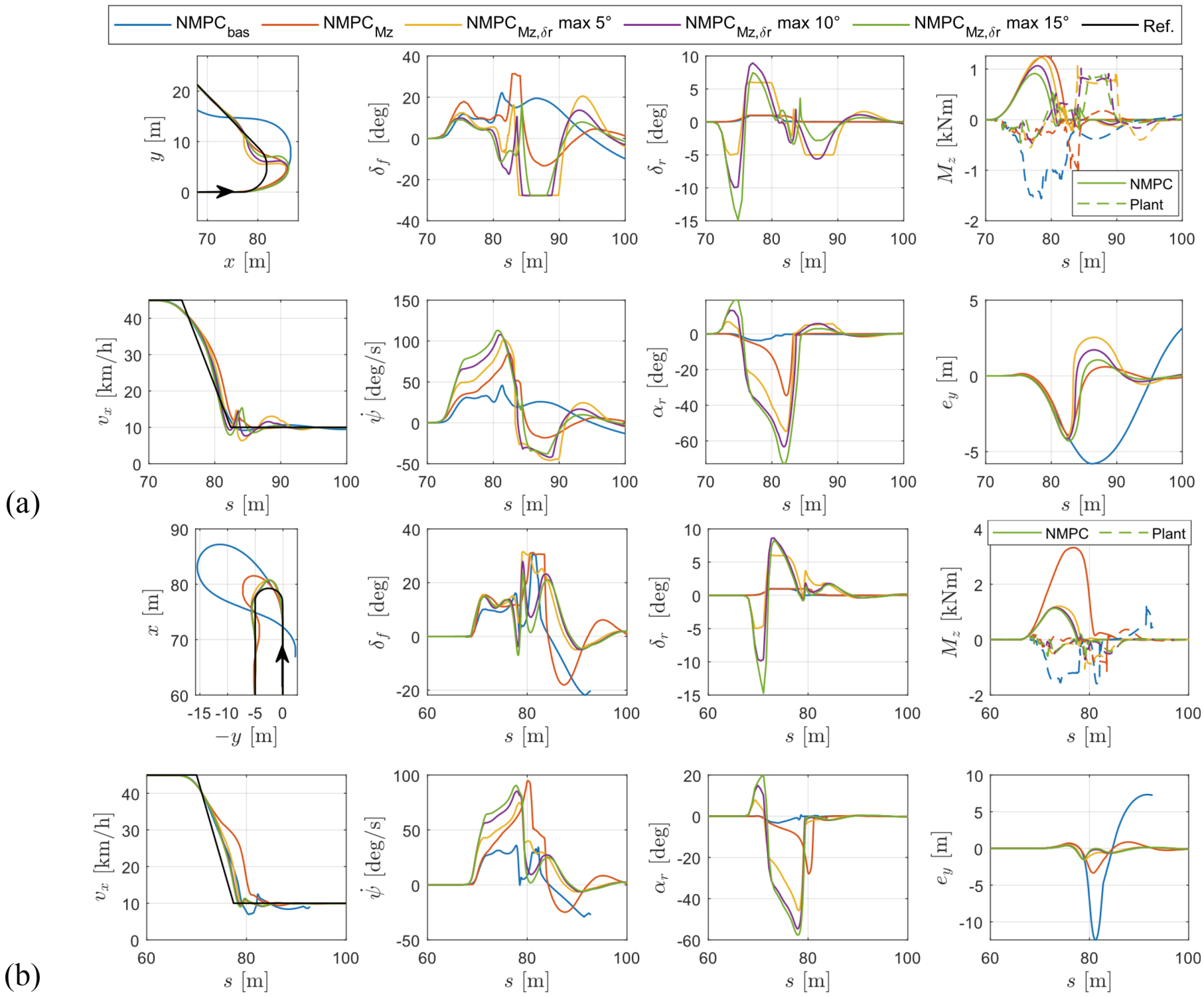

**Fig. 2.** Simulation results for (a) the 135-deg turn and (b) the U-turn, for $NMPC_{bas}$, $NMPC_{M_z}$, and $NMPC_{M_z,\delta_r}$ with maximum RWS angle constraints equal to 5, 10, and 15 deg.

**Table 1.** KPIs along the 135 deg-turn and U-turn (the best values are highlighted in bold).

| Case | Configuration | $\lvert e_y \rvert_{max}$ | $RMS_{e_y}$ | $RMS_{e_{v_x}}$ | $\lvert \beta \rvert_{max}$ |
|---|---|---|---|---|---|
| | | [m] | [m] | [km/h] | [deg] |
| 135-deg turn | $NMPC_{M_z}$ | **3.88** | **0.988** | 1.574 | 17.3 |
| | $NMPC_{M_z,\delta_r}$, $\delta_{r,max} = 5$ deg | 4.09 | 1.227 | 1.553 | 38.6 |
| | $NMPC_{M_z,\delta_r}$, $\delta_{r,max} = 10$ deg | 4.16 | 1.131 | **1.176** | 53.1 |
| | $NMPC_{M_z,\delta_r}$, $\delta_{r,max} = 15$ deg | 4.30 | 1.139 | 1.211 | 63.0 |
| U-turn | $NMPC_{M_z}$ | 3.347 | 0.945 | 3.265 | 16.9 |
| | $NMPC_{M_z,\delta_r}$, $\delta_{r,max} = 5$ deg | 1.572 | 0.397 | 1.618 | 27.8 |
| | $NMPC_{M_z,\delta_r}$, $\delta_{r,max} = 10$ deg | **1.476** | 0.337 | 1.223 | 39.6 |
| | $NMPC_{M_z,\delta_r}$, $\delta_{r,max} = 15$ deg | 1.493 | **0.329** | **1.110** | 44.2 |

## 4 Conclusion

The study introduced a nonlinear model predictive controller ($NMPC_{M_z,\delta_r}$) for the concurrent actuation of the front and rear steering angles, the total longitudinal tyre force and its distribution between the axles, as well as the direct yaw moment, to execute drifting manoeuvres when beneficial to the path tracking (PT) performance in emergency conditions. The simulation results, based on an experimentally validated vehicle model, highlight that: i) drifting is very effective in dealing with challenging trajectories that PT controllers coupled with conventional vehicle stability controllers cannot achieve (see the $NMPC_{bas}$ results), thereby enhancing the collision avoidance capability; and ii) in the more demanding scenario, i.e., the U-turn test, the rear steering actuation of $NMPC_{M_z,\delta_r}$ significantly enhances the tracking performance, by reducing the maximum lateral error by ~1.8 m compared to $NMPC_{M_z}$.